\documentclass[fleqn]{tlp}

\usepackage{latexsym}
\usepackage{amssymb}
\usepackage{amsmath}
\usepackage[all]{xy}
\usepackage{pdfpages} 

\newcommand{\ignore}[1]{}

\newtheorem{definition}{Definition} 
\newtheorem{example}{Example} 

\makeatletter
\newenvironment{prog}{\vspace{0.7ex}\par
\setlength{\parindent}{0.7cm}
\obeylines\@vobeyspaces\tt}{\vspace{0.7ex}\noindent
}
\makeatother
\newcommand{\startprog}{\begin{prog}}
\newcommand{\stopprog}{\end{prog}\noindent}

\newcommand{\ren}{\mathit{ren}}

\newcommand{\query}{\mathit{query}}
\newcommand{\mapg}{\stackrel{gr}{\mapsto}}
\newcommand{\maps}{\stackrel{sh}{\mapsto}}

\def\defemb#1#2{\expandafter\def\csname #1\endcsname
                              {\relax\ifmmode #2\else\hbox{$#2$}\fi}}
\defemb{cA}{{\cal A}}
\defemb{cB}{{\cal B}}
\defemb{cC}{{\cal C}}
\defemb{cD}{{\cal D}}
\defemb{cE}{{\cal E}}
\defemb{cF}{{\cal F}}
\defemb{cG}{{\cal G}}
\defemb{cH}{{\cal H}}
\defemb{cI}{{\cal I}}
\defemb{cJ}{{\cal J}}
\defemb{cL}{{\cal L}}
\defemb{cM}{{\cal M}}
\defemb{cN}{{\cal N}}
\defemb{cO}{{\cal O}}
\defemb{cP}{{\cal P}}
\defemb{cQ}{{\cal Q}}
\defemb{cR}{{\cal R}}
\defemb{cS}{{\cal S}}
\defemb{cT}{{\cal T}}
\defemb{cU}{{\cal U}}
\defemb{cV}{{\cal V}}
\defemb{cX}{{\cal X}}
\defemb{cZ}{{\cal Z}}

\newcommand{\var}{{\cV}ar}

\newcommand{\nil}{[\:]}

\renewcommand{\phi}{\varphi}

\newcommand{\pred}{\mathit{pred}}

\def \tuple#1{\langle #1 \rangle}

\long\def\comment#1{}

\begin{document}

\title[Annotation of Logic Programs for Independent AND-Parallelism]%
{Annotation of Logic Programs for Independent AND-Parallelism 
by Partial Evaluation%
\thanks{This work has been partially supported by the Spanish
  \emph{Ministerio de Econom\'{\i}a y Competitividad (Secretar\'{\i}a
    de Estado de Investigaci\'on, Desarrollo e Innovaci\'on)} under
  grant TIN2008-06622-C03-02 and by the \emph{Generalitat Valenciana}
  under grant PROMETEO/2011/052.}  }

\author[Germ\'an Vidal]
{GERMAN VIDAL \\
MiST, DSIC,
Universitat Polit\`ecnica de Val\`encia\\
Camino de Vera, S/N, 46022 Valencia, Spain    \\
E-mail: gvidal@dsic.upv.es
}

\pagerange{\pageref{firstpage}--\pageref{lastpage}}
\volume{\textbf{10} (3):}
\jdate{March 2002}
\setcounter{page}{1}
\pubyear{2002}

\maketitle

\label{firstpage}

\begin{abstract}
  Traditional approaches to automatic AND-parallelization of logic
  programs rely on some static analysis to identify independent goals
  that can be safely and efficiently run in parallel in any possible
  execution. In this paper, we present a novel technique for
  generating annotations for independent AND-parallelism that is based
  on partial evaluation. Basically, we augment a simple partial
  evaluation procedure with (run-time) groundness and variable sharing
  information so that parallel conjunctions are added to the residual
  clauses when the conditions for independence are met. In contrast to
  previous approaches, our partial evaluator is able to transform the
  source program in order to expose more opportunities for
  parallelism. To the best of our knowledge, we present the first
  approach to a \emph{parallelizing} partial evaluator.

  \emph{To appear in Theory and Practice of Logic Programming.}
\end{abstract}


\section{Introduction} \label{intro}

With the widespread adoption of multi-core processors, the generation
of automatic parallelizing compilers becomes an urgent need. On the
other hand, there exist a number of program optimization techniques
(like partial evaluation \cite{JGS93}) that have not considered the
introduction of parallelism so far, thus limiting its potential for
improving program performance.

In this work, we tackle the definition of a parallelizing
\emph{partial evaluator} which is able to automatically generate
annotations for independent AND-parallelism from logic programs.  In
contrast to traditional approaches to automatic AND-parallelization of
logic programs (which rely on some static analyses to identify
independent goals that can be safely and efficiently run in parallel
in any possible execution), our approach combines both run-time
analyses and the dynamic information gathered during partial
evaluation. Furthermore, it allows us to transform the source program
in order to expose more opportunities for parallelism (e.g., we can
have different specializations of a given clause so that some of them
are parallelized and some are not, without adding run-time
conditions).

\paragraph{Partial evaluation.} \label{pe-sec}

Partial evaluation \cite{JGS93} is a well-known technique for program
specialization. 
From a broader perspective, some partial evaluators are also able to
optimize programs further by, e.g., shortening computations, removing
unnecessary data structures and composing several procedures or
functions into a comprehensive definition.  Within this broader
approach, given a program and a \emph{partial} (incomplete) call, the
essential components of partial evaluation are: the construction of a
\emph{finite} representation|generally a graph|of the possible
executions of (any instance of) the partial call, followed by the
systematic extraction of a \emph{residual} program (i.e., the
partially evaluated program) from this graph. Intuitively,
optimization can be achieved by compressing paths in the graph, by
deleting unfeasible paths, and by renaming expressions while removing
unnecessary function symbols.  
In this paper, we propose a novel source of optimization based on
transforming some sequential constructions of residual programs into
parallel ones.

The theoretical foundations of partial evaluation for (normal) logic
programs was first put on a solid basis by Lloyd and Shepherdson
\citeyear{LS91}.  When \emph{pure} logic programs are considered, the
term \emph{partial deduction} is often used. Roughly speaking, in
order to compute the partial deduction of a logic program $P$ w.r.t.\
a set of atoms $\cA = \{A_1,\ldots,A_n\}$, one should construct
finite|possibly incomplete|SLD trees for the atomic goals $\leftarrow
A_1$, \ldots, $\leftarrow A_n$, such that every leaf is either
successful, a failure, or only contains atoms that are
\emph{instances} of $\{A_1,\ldots,A_n\}$; this is the so-called
\emph{closedness} condition \cite{LS91}. The residual program then
includes a \emph{resultant} of the form $A_i\sigma \leftarrow Q$ for
every non-failing root-to-leaf derivation $\leftarrow A_i
\hookrightarrow^\ast_\sigma\; \leftarrow Q$ in the SLD trees. Similarly,
we say that a residual program $P'$ is \emph{closed} when every atom
in the body of the clauses of $P'$ is an instance of a partially
evaluated atom (i.e., an appropriate specialized definition exists).

From an algorithmic perspective, in order to partially evaluate a
program $P$ w.r.t.\ an atom $A$, one starts with the initial set
$\cA_1 = \{A\}$ and builds a \emph{finite} (possibly incomplete) SLD
tree for $\leftarrow A$.  Then, all atoms in the leaves of this SLD
tree which are not instances of $A$ are added to the set, thus
obtaining $\cA_2$, and so forth.  In order to keep the sequence
$\cA_1,\cA_2,\ldots$ finite, some \emph{generalization} is often
required, e.g., by replacing some predicate arguments with fresh
variables. Some variant of the \emph{homeomorphic embedding} ordering
\cite{Leu02} is often used to detect potential sources of
non-termination.

A sketch of this algorithm is shown in Figure~\ref{pdalg}, where the
unfolding rule $\mathit{unf}(\cA_i)$ builds finite SLD trees for the
atoms in $\cA_i$ and returns the associated resultants, function
$\mathit{atoms}$ returns the atoms in the bodies of these resultants,
and the abstraction operator $\mathit{abs}(\cA_i,\cA')$ returns an
approximation of $\cA_i\cup\cA'$ so that the sequence
$\cA_1,\cA_2,\ldots$ is kept finite.

\begin{figure}[h]
\begin{center}
  \rule{\linewidth}{1pt}\\[1ex]
  \begin{minipage}{0.45\linewidth}
    \begin{description}
    \item[Initialization:] $i := 1$;~ $\cA_i := \{ A \};$
    \item[Repeat]
    \item[] \hspace{5ex}$\cA_{i+1} :=
      \mathit{abs}(\cA_i,\mathit{atoms}(\mathit{unf}(\cA_i)))$;
        \item[] \hspace{5ex}$i := i+1$ 
    \item[Until] $\cA_i \approx \cA_{i-1}$  (variants)
    \item[Return] $\mathit{unf}(\cA_i)$
    \end{description}
  \end{minipage}
  \\[1ex]
  \rule{\linewidth}{1pt}
\end{center}
\caption{Partial evaluation procedure} \label{pdalg}
\end{figure}

\paragraph{Motivation.}

Depending on \emph{when} control issues|like deciding which atoms
should or should not be unfolded|are addressed, two main approaches
to partial evaluation can be distinguished. In \emph{offline}
approaches to partial evaluation, these decisions are taken beforehand
by means of a so called \emph{binding-time} analysis (where we know
which parameters are known but not their values).  In contrast,
\emph{online} partial evaluators take decisions on the fly (so that
actual values of static data are available).

While offline partial evaluators are usually faster, online ones
produce more accurate results. Partial evaluators for logic programs
have mostly followed the online approach (e.g., SAGE \cite{Gur94},
Mixtus \cite{Sah90}, SP \cite{Gal91}, ECCE \cite{LEVCF06}), though
some offline partial evaluators have been also developed (e.g., LOGEN
\cite{LEVCF06}).

Recently, we have proposed in \cite{Vid10} a hybrid approach to
partial evaluation that does not fit well in neither the offline nor
the online style of partial evaluation. Basically, we follow a typical
online partial evaluation scheme, but augment it with run-time
information gathered from a pre-processing static analysis. There are
some previous approaches that combine the online and offline styles of
partial evaluation. However, the novelty is that \cite{Vid10}
considers collecting \emph{run-time} information rather than
\emph{partial evaluation} time information in a pre-processing stage
(as \emph{binding-time} analyses do). 

In this paper, we want to push this approach forward by defining a
parallelizing partial evaluator that generates annotations for
independent AND-parallelism.

As it is well known, two goals $(G_1,G_2)\theta$ are \emph{strictly
  independent} if, for every pair of variables $(x,y) \in
(\var(G_1),\var(G_2))$, either (i) they are equal, $x=y$, and
$x\theta$ is ground (i.e., $\var(x\theta)=\emptyset$) or (ii) they are
different, $x\neq y$, and their values, $x\theta$ and $y\theta$, do
not share a common variable (i.e.,
$\var(x\theta)\cap\var(y\theta)=\emptyset$).
In order to have this information available at partial evaluation
time, we need some \emph{run-time} information that is not usually
present in partial evaluation schemes. For this purpose, we introduce
a hybrid partial evaluation scheme with the following features:
\begin{itemize}
\item First, a pre-processing stage performs both a groundness and
  sharing analysis, so that we get call and success patterns for each
  predicate.
\item Then, we apply a rather simple partial evaluation stage that
  only performs one-step unfolding. This is very limited in general
  and propagates almost no information. However, in our context, we do
  not aim at aggressively propagating static data but only groundness
  and sharing information. In this way, the potential for generating
  annotations for the implicit independent AND-parallelism can be
  better evaluated.
\item Finally, a post-processing stage extracts residual rules from
  the partial evaluation computations and, in some cases, replaces
  sequential conjunctions by parallel ones, thus boosting the
  performance of the residual program.
\end{itemize}
A proof-of-concept implementation of the parallelizing partial
evaluator is available at
\verb$http://kaz.dsic.upv.es/litep.html$. Despite its simplicity (a
thousand lines of Prolog code), the results for definite logic
programs (including some built-in's) are very encouraging.

The paper is organized as follows. Section~\ref{ppeval} presents the
different stages of our parallelizing partial evaluation scheme. Then,
Section~\ref{experiments} summarizes our findings from an experimental
evaluation of the new technique and, finally, Sect.~\ref{future}
concludes and discusses some possibilities for future work.
Correctness results can be found in the online appendix.

\section{Parallelizing Partial Evaluation} \label{ppeval}

In this section, we present our partial evaluation scheme in a
stepwise manner. We do so for clarity of presentation but these stages
can be interleaved (and actually they are in our implementation).

\subsection{Pre-Processing Stage} \label{pre-section}

Our pre-processing stage consists of two different analyses. The first
one is a simple call and success pattern analysis that resembles a
mode analysis. The formal definition of the analysis can be found
elsewhere (e.g., in \cite{LV08}).

We consider \emph{groundness} call and success patterns $\pi$ denoted
by a list of natural numbers which represent the (definitively) ground
arguments of a predicate.  The underlying abstract domain is thus very
simple: $\{$\textsf{definitively ground},~\textsf{possibly
  non-ground}$\}$.  As mentioned in \cite{LV08}, the analysis could be
made more precise by considering a richer abstract domain (including
elements like \textsf{list}, \textsf{nonvar}, etc).  This is
orthogonal to the topics of this paper and thus we keep the two
element domain for simplicity.
The greatest lower bound operator $\sqcap$ on patterns is defined in
the natural way by the set union, i.e., given two patterns
$\pi_1,\pi_2$ for predicate $p/n$, we let $\pi_1\sqcap\pi_2 = \pi_1
\cup \pi_2$.

Basically, given an initial query and the groundness call patterns for
the atoms in this query, the analysis infers for every predicate $p/n$
a number of call and success patterns of the form $p/n: \pi_{in}
\mapg \pi_{out}$ such that $\pi_{in}$ and $\pi_{out}$ are subsets of
$\{1,\ldots,n\}$ denoting the arguments $\pi_{out}$ of $p/n$ which are
definitely ground after a successful derivation, assuming that it is
called with ground arguments $\pi_{in}$.
The analysis is started with a number of
\emph{entry points} to the program, together with their initial
groundness call patterns.

\begin{example} \label{app1}
  Consider the well known definition of $append/3$:
  \[
  \begin{array}{lll}
    append(\nil,Y,Y).\\
    append([H|T],Y,[H|TY]) & \leftarrow & append(T,Y,TY).
  \end{array}
  \]
  Given the initial groundness call patterns $\pi_1 = \{1\}$ and
  $\pi_2 = \{1,2\}$ for $append/3$, the call and success pattern
  analysis would return the following mappings:
  \[
  \begin{array}{lllllllll}
  append/3:  & \{1\} & \mapg & \{1\} & & 
  append/3:  & \{1,2\} & \mapg & \{1,2,3\} \\
  \end{array}
  \]
  Their meaning should be clear: if $append(t_1,t_2,t_3)$ is called
  with $t_1$ ground, we can only ensure that $t_1$ will be ground
  after a successful derivation. In contrast, if it is called with
  both $t_1$ and $t_2$ ground, then $t_3$ will be also ground after a
  successful derivation.
\end{example}
For guaranteeing the independence of goals, we also consider the
information gathered by a dependency analysis like that of
\cite{Deb89}.
Basically, for a given predicate $p/3$, the analysis computes mappings
with \emph{sharing} call and success patterns $\mu$ like, e.g.,
$\tuple{\{1,2\},\{1,2,3\},\{2,3\}}$, which indicates that the first
argument may share variables with the second argument, the second
argument may share variables with the first and third arguments, and
the third argument may share variables with the second
argument. Again, the analysis infers for every predicate $p/n$ a
number of call and success patterns of the form $p/n: \mu_{in} \maps
\mu_{out}$ such that $\mu_{in}$ and $\mu_{out}$ belong to the domain
$2^{\{1,\ldots,n\}}\times\ldots\times2^{\{1,\ldots,n\}}$ (a tuple of
$n$ sets) and $\mu_{out}$ denotes the dependencies of $p/n$ which hold
after a successful derivation, assuming that it is called with the
dependencies denoted by $\mu_{in}$.\footnote{We note that a sharing
  pattern like $\tuple{\{1\},\{2\},\{3\}}$ assumes that all three
  argument are independent and, moreover, that no variable sharing can
  be introduced through a single argument; i.e., we assume that
  predicate arguments are always linear. We keep this restriction for
  simplicity but could easily be overcome.
}

In this case, the least upper bound operator $\sqcup$ on sharing
patterns is defined as follows: given patterns $\mu =
\tuple{\vartheta_1,\ldots,\vartheta_n}$ and
$\mu'=\tuple{\vartheta'_1,\ldots,\vartheta'_n}$ for some predicate
$p/n$, we have $\mu\sqcup\mu'=\tuple{\vartheta_1 \cup \vartheta'_1,
  \ldots, \vartheta_n \cup \vartheta'_n}$. 
Note that, in contrast to the greatest lower bound on groundness
patterns that may increase the number of ground variables (and thus
the accuracy of the result), the least upper bound on sharing
patterns may lose accuracy since more dependencies can be obtained.

\begin{example} \label{app2} Consider again $append/3$. Given the
  sharing call patterns $\mu_1=\tuple{\{1\},\{2\},\{3\}}$ and
  $\mu_2=\tuple{\{1,2\},\{1,2\},\{3\}}$, the dependency analysis would
  return the following:
  \[
  \begin{array}{llll}
  append/3:  & \tuple{\{1\},\{2\},\{3\}} & \maps & \tuple{\{1,3\},\{2,3\},\{1,2,3\}} \\
  append/3:  & \tuple{\{1,2\},\{1,2\},\{3\}} & \maps & \tuple{\{1,2,3\},\{1,2,3\},\{1,2,3\}} \\
  \end{array}
  \]
  Here, we consider two possibilities: first, if $append$ is called
  with three independent arguments then, after a successful
  derivation, the third argument may be bound to a value that shares
  variables with either the first and the second arguments; on the
  other hand, if $append$ is called with the two first arguments bound
  to terms containing shared variables, then all three arguments may
  depend on each other after a successful derivation.
\end{example}
%

\subsection{Partial Evaluation Stage} \label{pe-section}

Now, we present the proper partial evaluation stage of the
parallelizing partial evaluator.

In principle, one could consider checking independence of goals using
the information available solely at partial evaluation time. This
approach, however, would be generally incorrect for a number of
reasons. First, the notion of \emph{closedness} (see
Sect.~\ref{pe-sec}) allows run-time atoms to be covered by instances
of partial evaluation atoms. Therefore, $q(X,X)$ is closed w.r.t.\
$q(X,Y)$. This means that goals can be independent at partial
evaluation time but need not be independent at run-time. Moreover,
whenever we split a goal of an incomplete computation into atomic
subgoals, we are also loosing some \emph{context} information that
might be essential for checking independence, as the following example
illustrates:

\begin{example}
  Consider the following program
  \[
  \begin{array}{l}
    p(X,Y) ~ \leftarrow ~ q(X),r(Y).\\
    eq(X,X).  \\
    \ldots \\
  \end{array}
  \]
  Given the goal $eq(A,B),p(A,B)$, if we split it into its atomic
  subgoals $eq(A,B)$ and $p(A,B)$, and partially evaluate them
  independently, we could derive the goal $q(A),r(B)$ and
  \emph{incorrectly} assume that $q(A)$ and $r(B)$ are independent.
\end{example}
Furthermore, the use of an abstraction operator might also involve the
loss of some dependencies (e.g., generalizing $p(X,Y,f(Y))$ to
$p(X,Y,Z)$ with $Z$ a fresh variable).

In summary, the information available at partial evaluation time is
not enough to determine the run-time independence of a
goal.\footnote{Of course, we could avoid splitting goals, do not use
  an abstraction operator and only allow variants to be closed, but
  then the termination of partial evaluation could not be ensured.}
Therefore, as mentioned before, in this paper we consider that the
partial evaluator includes a pre-processing stage where
\emph{run-time} groundness and sharing information is gathered.

In particular, we design a rather simple partial evaluator with the
following distinguishing features:
\begin{itemize}
\item only one-step unfolding of atomic goals is performed;
\item no static data are provided (i.e., the initial goal has
  different variables as arguments);
\item every atomic goal is enriched with groundness and sharing call
  patterns that are propagated through partial evaluation.
\end{itemize}
The fact that we do not consider partially instantiated initial goals,
together with the fact that only one-step unfolding is performed,
allows us to better identify the potential for generating annotations
for independent AND-parallelism. Moreover, it makes the online partial
evaluator scale up better to medium and large applications.

Our partial evaluator deals with sets of \emph{extended} atoms
(instead of sets of atoms, as in the algorithm of Figure~\ref{pdalg}).

\begin{definition}[extended atom]
  We consider extended atoms of the form $(A,\pi,\mu)$ where $A$ is an
  atom, $\pi$ is a groundness call pattern for $A$, and $\mu$ is a
  sharing call pattern for $A$. This notion is extended in the natural
  way to queries and goals. We denote the empty extended query by
  $true$.

  Given an extended query $\cQ$, we introduce the following auxiliary
  function:\linebreak $\query(\cQ) = A_1,\ldots,A_n$, if $\cQ =
  (A_1,\pi_1,\mu_1),\ldots,(A_n,\pi_n,\mu_n)$.
\end{definition}
The number of different specialized versions of an atom will be
determined, not only by its shape (as it is usually the case), but
also by the different combinations of groundness and sharing call
patterns. For instance, $(p(X,Y),\{1,2\},\tuple{\{1\},\{2\}})$ and
$(p(X,Y),\{1\},\tuple{\{1\},\{2\}})$ would give rise to different
specialized versions.

Another distinguishing feature of our scheme is that, in contrast to
previous approaches, we do not explicitly distinguish between the
so-called \emph{local} and \emph{global} levels (as in
\cite{Gal93}). Rather, we construct a single partial evaluation tree
that comprises both levels. Moreover, our partial evaluation process
performs just one pass since residual rules can be produced
immediately after every unfolding step (rather than in a post-process,
as it is often done since the unfolding tree can be modified during
the partial evaluation process).

In the following, we denote by $\pi(A)$ the (definite) \emph{ground}
arguments of $A$ according to $\pi$, i.e., $\pi(p(s_1,\ldots,s_k)) =
\{ s_j \mid j\in\pi\}$. Also, we denote by $\mu(A)$ the set of
(possibly) shared variables in $A$ according to $\mu$, i.e.,
$\mu(p(s_1,\ldots,s_k)) = \{ (x,y) \in (\var(s_i),\var(s_j)) \mid
i,j\in\{1,\ldots,k\}, i\neq j, \{i,j\}\subseteq s \in \mu\}$.
Before introducing the notion of SLD resolution over extended queries,
we need the following preparatory definition, which is used to
propagate groundness and sharing call patterns to the atoms in the
body of a clause.

\begin{definition}[entry procedure] \label{entrydef} Let $H \leftarrow
  B_1,\ldots,B_n$ be a clause and $(A,\pi,\mu)$ an extended atom such
  that $A$ and $H$ unify. We denote with $entry$ a function that
  propagates $\pi$ and $\mu$ to $B_1,\ldots,B_n$. Formally,
  $entry(\pi,\mu,(H\leftarrow B_1,\ldots,B_n)) =
  ((B_1,\pi_1,\mu_1),\ldots,(B_n,\pi_n,\mu_n))$ if, for all
  $B_i=p_i(t_{i1},\ldots,t_{im_i})$, $i=1,\ldots,n$, the following
  conditions hold:
    \begin{itemize}
    \item $j \in \pi_i$ iff $\var(t_{ij})\subseteq\var(\pi(H))$ (i.e.,
      all variables in $t_{ij}$ are ground in $H$ according to $\pi$).

    \item $\{1,\ldots,m_i\}\supseteq\{j_1,\ldots,j_k\}\in\mu_i$ iff
      there are (non necessarily different) variables
      $(x_{j_1},\ldots,x_{j_k}) \in
      (\var(t_{ij_1}),\ldots,\var(t_{ij_k}))$ such that for every pair
      of different variables $x_{j_r},x_{j_s}$, we have
      $(x_{j_r},x_{j_s})\in\mu(H)$ (i.e., either the terms share some
      variable or have different variables that are shared in $H$
      according to $\mu$).
    \end{itemize}
\end{definition}
Note that the entry procedure is independent of $A$ (only its
associated groundness and sharing call patterns matter), since we want
the results for a partial evaluation time atom $A$ be valid for every
run-time atom $A\theta$.

\pagebreak 

\begin{example} \label{fib}
  Let us consider the following program for computing Fibonacci
  numbers:
  \[
  \begin{array}{llll@{~~~~~~~~~~~~~~~~~}l}
    (C_1) & fibonacci(0,1). && 
    (C_2) ~~ fibonacci(1,1). \\
    (C_3) & fibonacci(M,N) & \leftarrow &
    M > 1,~M1 ~is~ M-1,~fibonacci(M1,N1), \\
    &&& M2 ~is~ M-2,~fibonacci(M2,N2),~N ~is~ N1 + N2.  \\
  \end{array}
  \]
  Here, $entry(
  \{1\},\tuple{\{1\},\{2\}},C_3)$ returns
  the following extended query:
  \[
  \begin{array}{r@{~}c@{~}l}
    (M > 1, & \{1,2\}, & \tuple{\{1\},\{2\}}), \\
    (M1 ~is~ M-1, & \{2\}, & \tuple{\{1\},\{2\}}), \\
    (fibonacci(M1,N1), & \{\},& \tuple{\{1\},\{2\}}), \\
    (M2 ~is~ M-2, & \{2\}, & \tuple{\{1\},\{2\}}), \\
    (fibonacci(M2,N2), & \{\},& \tuple{\{1\},\{2\}}), \\
    (N ~is~ N1+N2, & \{\}, & \tuple{\{1\},\{2\}}) \\
  \end{array}
  \]
\end{example}
We are now ready to introduce the notion of extended SLD resolution:

\begin{definition}[extended SLD resolution]
  Extended SLD resolution, denoted by $\leadsto$, is a natural
  extension of SLD resolution over extended queries.  Formally, given
  a program $P$, an extended query $\cQ=
  (A_1,\pi_1,\mu_1),\ldots,(A_n,\pi_n,\mu_n)$, and a computation rule
  $\cR$, we say that $\leftarrow \cQ \leadsto_{P,\cR,\sigma}\:
  \leftarrow \cQ'$ is an \emph{extended SLD resolution step} for $\cQ$
  with $P$ and $\cR$ if the following conditions hold:\footnote{We
    often omit $P$, $\cR$ and/or $\sigma$ in the notation of an
    extended SLD resolution step when they are clear from the
    context.}
  \begin{itemize}
  \item $\cR(\cQ) = (A_i,\pi_i,\mu_i)$, $1\leq i\leq n$, is the selected
    extended atom,
  \item $H \leftarrow B_1,\ldots,B_m$ is a renamed apart clause of
    $P$, 
  \item $A_i$ and $H$ unify with $\sigma = mgu(A_i,H)$, and
  \item $\cQ' = entry(
    \pi_i,\mu_i,(H \leftarrow
    B_1,\ldots,B_m))\sigma$.\footnote{We let
      $((B_1,\pi_1,\mu_1),\ldots,(B_n,\pi_n,\mu_n))\sigma =
      (B_1\sigma,\pi_1,\mu_1),\ldots,(B_n\sigma,\pi_n,\mu_n)$.}
  \end{itemize}
\end{definition}
Trivially, extended SLD resolution is a conservative extension of SLD
resolution: given extended queries $\cQ,\cQ'$, we have that $\leftarrow \cQ
\leadsto_\sigma\:\leftarrow \cQ'$ implies $\leftarrow \query(\cQ)
\hookrightarrow_\sigma\: \leftarrow \query(\cQ')$.

In the following, we use $\pred(A)$ to denote the predicate symbol of
atom $A$.

As it is common practice, we avoid infinite unfolding by means of a
well-known strategy based on the use of the \emph{homeomorphic
  embedding} ordering. 
Intuitively, we say that atom $A_i$ \emph{embeds} atom $A_j$, denoted
by $A_i \unrhd A_j$, when $A_j$ can be obtained from $A_i$ by deleting
symbols (see \cite{Leu02} for a precise definition).

\begin{definition}[variant, embedding]
  We say that two (extended) atoms $(A,\pi,\mu)$ and $(A',\pi',\mu')$
  are \emph{variants}, denoted by $(A,\pi,\mu) \approx (A',\pi',\mu')$
  if there is a renaming substitution $\rho$ such that $A\rho= A'$,
  $\pi = \pi'$ and $\mu=\mu'$.
  
  We say that $(A,\pi,\mu)$ \emph{embeds} $(A',\pi',\mu')$, denoted by
  $(A,\pi,\mu) \trianglerighteq (A',\pi',\mu')$, if $A\trianglerighteq
  A'$, $\pi = \pi'$ and $\mu=\mu'$.
\end{definition}
Our partial evaluation semantics is formalized by means of the
(labelled) state transition system shown in Figure~\ref{ecpd}. The
partial evaluator deals with \emph{states}, defined as follows:

\begin{figure}[t]
  \rule{\linewidth}{1pt}
    \[ 
    \begin{array}{r@{~~~~~}c}

      \mathsf{(variant)} & {\displaystyle 
        \frac{\exists (A',\pi',\mu')\in memo.~ (A,\pi,\mu)\approx (A',\pi',\mu')} 
        {\tuple{(A,\pi,\mu),\cQ;memo} \stackrel{\textit{v}}{\longrightarrow} \tuple{\cQ;memo}}}
       \\[4ex]
      
      \mathsf{(failure)} & {\displaystyle 
        \frac{\not\exists \cQ'.~  \leftarrow (A,\pi,\mu) \leadsto_\sigma \: \leftarrow \cQ'} 
        {\tuple{(A,\pi,\mu),\cQ;memo} \stackrel{\textit{f}}{\longrightarrow} \tuple{\cQ;memo}}}
       \\[4ex]
      
       \mathsf{(embedding)} & {\displaystyle 
         \frac{\exists (A',\pi',\mu')\in memo.~ (A,\pi,\mu)\trianglerighteq (A',\pi',\mu') } 
         {\tuple{(A,\pi,\mu),\cQ;memo} \stackrel{\textit{e}}{\longrightarrow} \tuple{\cQ;memo}}}
       \\[4ex]
      
       \mathsf{(nonuser)} & {\displaystyle 
         \frac{ \mbox{$\pred(A)$ is not defined in the user's program clauses}} 
         {\tuple{(A,\pi,\mu),\cQ;memo} \stackrel{\textit{n}}{\longrightarrow} \tuple{\cQ;memo}}}
       \\[4ex]
      
       \mathsf{(parallel)} & {\displaystyle 
         \frac{ \leftarrow (A,\pi,\mu) \leadsto_\sigma \leftarrow \cQ' ~\wedge~ 
           \exists (\cQ_1,\cQ_2,\cQ_3,\cQ_4)\in\textit{partition}_\mu(\cQ')} 
         {\tuple{(A,\pi,\mu),\cQ;memo} \stackrel{\textit{p}}{\longrightarrow}_\sigma 
           \tuple{\cQ_1,\cQ_2,\cQ_3,\cQ_4,\cQ;memo\cup\{(A,\pi,\mu)\}}}}
       \\[4ex]
      
       \mathsf{(unfolding)} & {\displaystyle 
         \frac{ \leftarrow (A,\pi,\mu) \leadsto_\sigma \leftarrow \cQ' ~\wedge~ 
           \not\exists (\cQ_1,\cQ_2,\cQ_3,\cQ_4)\in\textit{partition}_\mu(\cQ')} 
         {\tuple{(A,\pi,\mu),\cQ;memo} \stackrel{\textit{u}}{\longrightarrow}_\sigma 
           \tuple{\textit{prop}(\cQ',true),\cQ;memo\cup\{(A,\pi,\mu)\}}}}
       \\
    \end{array}
    \]
  \rule{\linewidth}{1pt}
  \caption{Partial evaluation semantics} \label{ecpd}
\end{figure}

\begin{definition}[state]
  A \emph{state} is a pair of the form $\tuple{\cQ;memo}$ where $\cQ$
  is a sequence of extended atoms\footnote{Note that this sequence is
    not an extended query. Rather, this is the queue of (extended)
    atomic goals to be partially evaluated.} and $memo$ is a set of
  extended atoms (the atoms already partially evaluated, which are
  recorded to guarantee termination).

  An \emph{initial state} has the form $\tuple{(A,\pi,\mu);\{\}}$. A
  \emph{final state} has the form $\tuple{\epsilon;memo}$, where
  $\epsilon$ denotes an empty sequence.
\end{definition}
A \emph{successful} partial evaluation starts with an initial state
and (non-deterministically, because of the \textsf{unfolding} rule)
constructs a number of derivations of the form
$\tuple{(A,\pi,\mu);\{\}} \longrightarrow^\ast \tuple{\epsilon;\_}$, where
$\longrightarrow^\ast$ denotes the reflexive and transitive closure of
$\longrightarrow$. The process does not return anything but the trace
itself, that will be used for producing residual rules (see the next
section).

Let us now explain the rules of the partial evaluation semantics.
Rule (\textsf{variant}) discards an extended atom if it is a variant
of an already partially evaluated extended atom. Rule
(\textsf{failure}) also discards an extended atom when it cannot be
unfolded (e.g., when $A$ does not unify with the head of any clause).

The next rule, (\textsf{embedding}), discards an extended atom when it
embeds a previously partially evaluated extended atom. This rule is
necessary in order to ensure that partial evaluation always
terminates. Rule (\textsf{nonuser}) allows us to deal with built-in's
and other extra-logical features of Prolog by leaving calls to the
original predicates, as we will see in the next section.

The interesting rules are (\textsf{parallel}) and (\textsf{unfolding}). 
In the following, we assume a fixed left-to-right selection rule as in
Prolog. Therefore, we use a function $\mathit{prop}$ to propagate
groundness and sharing success patterns to the atoms to the right of a
given atom before splitting an extended query. This is necessary
because only the partial evaluation of atomic goals is allowed and,
thus, this information should be propagated before the query is split
into its constituents in order to avoid a serious loss of accuracy.

\begin{definition}[pattern propagation]
  Let $\cQ_1,\cQ_2$ be extended queries, with $\cQ_1 =
  (A_1,\pi_1,\mu_1),\ldots,(A_n,\pi_n,\mu_n)$ and $\cQ_2 =
  (A_{n+1},\pi_{n+1},\mu_{n+1}),\ldots,(A_m,\pi_m,\mu_m)$. We define
  the function $\mathit{prop}$ to propagate success patterns to the
  right as follows:\footnote{Note the non-standard use of function
    $entry$ to propagate success patterns to the right, despite the
    fact that $A_1\leftarrow A_2,\ldots,A_m$ is not really a program
    clause.}
  \begin{itemize}
  \item $\mathit{prop}(\cQ_1,\cQ_2) = \cQ_2$ if $n= 0$ (i.e.,
    $\cQ_1$ is an empty query);
  \item $\mathit{prop}(\cQ_1,\cQ_2) =
    ((A_1,\pi_1,\mu_1),\mathit{prop}(\cQ'_1,\cQ'_2))$ if $n>0$,\\
    $\pred(A_1):\pi_1\mapg\pi'_1$, $\pred(A_1):\mu_1\maps\mu'_1$,\\
    $\mathit{entry}(\pi'_1,\mu'_1,(A_1 \leftarrow A_2,\ldots,A_m)) =
    (A_2,\pi'_2,\mu'_2),\ldots,(A_m,\pi'_m,\mu'_m)$, \\
    $\cQ'_1 = (A_2,\pi_2\sqcap\pi'_2,\mu_2\sqcup\mu'_2),\ldots,
    (A_n,\pi_n\sqcap\pi'_n,\mu_n\sqcup\mu'_n)$, and \\ 
    $\cQ'_2 =
    (A_{n+1},\pi_{n+1}\sqcap\pi'_{n+1},\mu_{n+1}\sqcup\mu'_{n+1}),\ldots,
    (A_m,\pi_m\sqcap\pi'_m,\mu_m\sqcup\mu'_m)$.
  \end{itemize}
  Observe that the two arguments of function $prop$ are not needed for
  unfolding a goal. However, this formulation will become useful later
  when also using $\mathit{prop}$ to partition a goal.
\end{definition}

\begin{example} \label{exsld}
  Consider again the Fibonacci program of Example~\ref{fib} and the
  result of the $entry$ procedure. Thus we have $fibonacci(A,B)
  \leadsto_{\{A\mapsto M,B\mapsto N\}} (M > 1, \{1,2\},
  \tuple{\{1\},\{2\}})$, $(M1 ~is~ M-1, \{2\}, \tuple{\{1\},\{2\}}),
  (fibonacci(M1,N1), \{\}, \tuple{\{1\},\{2\}}), (M2 ~is~ M-2, \{2\},
  \tuple{\{1\},\{2\}}), (fibonacci(M2,N2), \{\},\tuple{\{1\},\{2\}}),
  (N ~is~ N1+N2, \{\}, \tuple{\{1\},\{2\}})$.
  We assume the following call and success patterns:
  \[
  \begin{array}{rllrll}
    is/2: & \{2\} \mapg \{1,2\} &&
    is/2: & \tuple{\{1\},\{2\}} \maps \tuple{\{1\},\{2\}} \\
    fibonacci/2: & \{1\} \mapg \{1,2\} &&
    fibonacci/2: & \tuple{\{1\},\{2\}} \maps \tuple{\{1\},\{2\}} \\
  \end{array}
  \]
  Then, for instance, we have
   \[
   \begin{array}{r@{}l@{}ll}
     \mathit{prop}( & ( & (M>1,\{1,2\},\tuple{\{1\},\{2\}}), ~
     M1 ~is~ M-1,\{2\},\tuple{\{1\},\{2\}}), \\
     && (fibonacci(M1,N1),\{\},\tuple{\{1\},\{2\}}), ~
      (M2 ~is~ M-2,\{2\}, \tuple{\{1\},\{2\}}),\\
     && (fibonacci(M2,N2), \{\},\tuple{\{1\},\{2\}}), 
      (N ~is~ N1+N2, \{\}, \tuple{\{1\},\{2\}})), ~true)  \\
     = & ( & (M>1,\{1,2\},\tuple{\{1\},\{2\}}),~
      (M1 ~is~ M-1,\{2\},\tuple{\{1\},\{1\}}), \\
     && (fibonacci(M1,N1),\{1\},\tuple{\{1\},\{2\}}),~ 
      (M2 ~is~ M-2,\{2\}, \tuple{\{1\},\{2\}}), \\
     && (fibonacci(M2,N2), \{1\},\tuple{\{1\},\{2\}}), ~
      (N ~is~ N1+N2, \{2\}, \tuple{\{1\},\{2\}})) \\
   \end{array}
   \]
   so we know that, when the last call $N ~is~ N1+N2$ is performed,
   $N1+N2$ is ground.
\end{example}
Before explaining the rules (\textsf{parallel}) and
(\textsf{unfolding}), we still need one more auxiliary function,
\textit{partition}, which is used to check if a query contains some
subgoals that can be executed in parallel (i.e., if they are
strictly independent at run-time):

\begin{definition}[partition]
  Let $(A,\pi,\mu)$ be an extended atom such that
  $(A,\pi,\mu)\leadsto_\sigma \cQ$. We introduce the function
  $\mathsf{partition}_\mu$ as follows:\footnote{In order not to
    encumber the notation, we assume that $\cQ'_i$ refers to the same
    extended query $\cQ_i$ after some processing.}
  \begin{itemize}
  \item $\mathit{partition}_\mu(\cQ) =
    (\cQ'_1,\cQ''_2,\cQ''_3,\cQ''''_4)$
    if $\cQ$ contains at least two extended atoms,\\
    $\cQ = \cQ_1,\cQ_2,\cQ_3,\cQ_4$, with $\cQ_2$ and $\cQ_3$ non-empty queries,\\
    $(\cQ'_1,(\cQ'_2,\cQ'_3,\cQ'_4)) = \mathit{prop}(\cQ_1,(\cQ_2,\cQ_3,\cQ_4))$,\\
    $\cQ'_2$ and $\cQ'_3$ are independent,
    \\
    $(\cQ''_2,\cQ''_4) = \mathit{prop}(\cQ'_2,\cQ'_4)$,
    $(\cQ''_3,\cQ'''_4) = \mathit{prop}(\cQ'_3,\cQ''_4)$, and
    $\cQ''''_4 = \mathit{prop}(\cQ'''_4,true).$
  \end{itemize}
  Here, strict independence of $\cQ'_2$ and $\cQ'_3$ is checked using
  the standard notion (see Section~\ref{intro}) and taking into
  account the groundness call patterns available from the extended
  atoms and the sharing call pattern for the head of the clause, i.e., the
  variables in $\var(\cQ'_2)\cap\var(\cQ'_3)$ must be ground according
  to the groundness call patterns in $\cQ'_2,\cQ'_3$ and each pair of
  different variables $(x,y)\in(\var(\cQ'_2),\var(\cQ'_3))$ should not
  be shared in the head of the clause according to $\mu$.
\end{definition}

\begin{example} \label{ex2} Consider again the Fibonacci program of
  Example~\ref{fib} and the extended SLD resolution step of
  Example~\ref{exsld}. By applying function \textit{partition} to the
  derived extended query, we get
  \[
  \begin{array}{l@{}lll}
    \cQ_1 = &  (M>1,\{1\},\tuple{\{1\},\{2\}}), \\
    \cQ_2 = & (M1 ~is~ M-1,\{2\},\tuple{\{1\},\{2\}}),
    (fibonacci(M1,N1),\{1\},\tuple{\{1\},\{2\}}), \\
    \cQ_3 = &  (M2 ~is~ M-2,\{2\}, \tuple{\{1\},\{2\}}),
    (fibonacci(M2,N2), \{1\},\tuple{\{1\},\{2\}}), \\
    \cQ_4 = & (N ~is~ N1+N2, \{2\}, \tuple{\{1\},\{2\}})\\
  \end{array}
  \]
  which means that the queries $(M1 ~is~ M-1,fibonacci(M1,N1))$ and $(M2
  ~is~ M-2,fibonacci(M2,N2))$ can be safely run in parallel at run-time.
\end{example} 
Now, rules (\textsf{parallel}) and (\textsf{unfolding}) should be
clear. When an atom is unfolded and the body of the selected clause
can be run in parallel (which is determined by function
\textit{partition}), rule (\textsf{parallel}) applies. Note that we
consider a simple algorithm where the atoms of a query cannot be
reordered (i.e., we respect Prolog's computation rule). Of course,
more elaborated strategies exist (see, e.g., \cite{MBBH99,GH09}), but
we consider them out of the scope of this paper.

When the body of the clause cannot be partitioned so that some
subgoals are run in parallel, rule (\textsf{unfolding}) applies (which
will give rise to a sequential clause, as we will see later). Here, we
apply function $\mathit{prop}$ in order to propagate groundness and
sharing information to the extended atoms before they are split in the
next step (since only the unfolding of atomic goals is considered).

In both rules, we add the selected extended atom to the set of already
partially evaluated extended atoms.

All transition rules are labelled with a letter that identifies the
rule applied. This will become useful to generate residual rules (see
the next section).

\begin{example} \label{ex3} Consider again the Fibonacci program of
  Example~\ref{fib}. Given the initial state
  \[
  \cS_0 = \tuple{(fibonacci(A,B),\{1\},\tuple{\{1\},\{2\}}),\{\}}
  \]
  we have three partial evaluation derivations starting from
  $\cS_0$:
  \[
  \begin{array}{l@{~}l}
    \cS_0 & \stackrel{u}{\longrightarrow}_{\{A\mapsto 0,B\mapsto 1\}} 
    \tuple{\epsilon,\{(fibonacci(A,B),\{1\},\tuple{\{1\},\{2\}})\}} \\
    \cS_0 & \stackrel{u}{\longrightarrow}_{\{A\mapsto 1,B\mapsto 1\}} 
    \tuple{\epsilon,\{(fibonacci(A,B),\{1\},\tuple{\{1\},\{2\}})\}} \\
    \cS_0 & \stackrel{p}{\longrightarrow}_{\{A\mapsto M,B\mapsto N\}} 
    \tuple{(\cQ_1,\cQ_2,\cQ_3,\cQ_4),\{(fibonacci(A,B),\{1\},\tuple{\{1\},\{2\}})\}} \\
    & \stackrel{n}{\longrightarrow} 
    \tuple{(\cQ_2,\cQ_3,\cQ_4),\{(fibonacci(A,B),\{1\},\tuple{\{1\},\{2\}})\}} \\
    & \stackrel{n}{\longrightarrow}
    \tuple{((fibonacci(M1,N1),\{1\},\tuple{\{1\},\{2\}}),\cQ_3,\cQ_4),\\
    & ~~~~~~ \{(fibonacci(A,B),\{1\},\tuple{\{1\},\{2\}})\}} \\
    & \stackrel{v}{\longrightarrow}
    \tuple{(\cQ_3,\cQ_4),
      \{(fibonacci(A,B),\{1\},\tuple{\{1\},\{2\}})\}} \\
    & \stackrel{n}{\longrightarrow}
    \tuple{((fibonacci(M2,N2),\{1\},\tuple{\{1\},\{2\}}),\cQ_4),\\
    & ~~~~~~ \{(fibonacci(A,B),\{1\},\tuple{\{1\},\{2\}})\}} \\
    & \stackrel{v}{\longrightarrow}
    \tuple{(\cQ_4),
      \{(fibonacci(A,B),\{1\},\tuple{\{1\},\{2\}})\}} \\
    & \stackrel{n}{\longrightarrow}
    \tuple{\epsilon,
      \{(fibonacci(A,B),\{1\},\tuple{\{1\},\{2\}})\}} \\
  \end{array}
  \]
  Note that predicates not defined in the user's program (like $>$ or
  $is$) are not unfoldable and that $\cQ_1,\cQ_2,\cQ_3,\cQ_4$ are the
  extended queries of Example~\ref{ex2}.
\end{example}

\subsection{Post-Processing Stage} \label{post-section}

Once the partial evaluation stage terminates, we produce renamed,
residual rules associated to the transitions of the partial evaluation
semantics. 
In the following, we assume that there is a function $\ren$ that takes
an extended atom and returns a renamed atom whose predicate name is
fresh and depends on the patterns of the extended atom. We do not
present the details of this renaming function here since it is a
standard renaming as introduced in, e.g., \cite{BL89,dSGJLMS99}. For
instance,
\[
\ren(\mathit{fibonacci}(X,Y),\{1\},\tuple{\{1\},\{2\}}) = 
\mathit{fibonacci}\_1\_1\_2(X,Y)
\]
Note, however, that non-user predicates are not renamed, e.g.,
\[
\ren(M1 ~is~ M-1,\{2\},\tuple{\{1\},\{2\}}) = 
M1 ~is~ M-1
\]
The generation of residual rules proceeds as follows:
\begin{itemize}
\item We do not generate residual clauses associated to the
  application of rules (\textsf{variant}) nor (\textsf{failure}).

\item For \textsf{embedding} steps of the form
  $\tuple{(A,\pi,\mu),\cQ;memo} \stackrel{\textit{e}}{\longrightarrow}
  \tuple{\cQ;memo}$ we produce a residual rule of the form
  $\ren(A,\pi,\mu) \leftarrow A$. This means that some atoms will not
  be closed but defined in terms of calls to the original predicates
  (and, thus, the clauses of the original program should be added to
  the residual program).

\item For \textsf{nonuser} steps $\tuple{(A,\pi,\mu),\cQ;memo}
  \stackrel{\textit{n}}{\longrightarrow} \tuple{\cQ;memo}$, we do not
  generate residual rules since non-user calls are not renamed.

\item For an \textsf{unfolding} step $\tuple{(A,\pi,\mu),\cQ;memo}
  \stackrel{\textit{u}}{\longrightarrow}_\sigma
  \tuple{\textit{prop}(\cQ',true),\cQ;memo\cup$ 
    $\{(A,\pi,\mu)\}}$, we produce a residual rule of the form
  \[
  \ren(A,\pi,\mu) \leftarrow
  \ren(B_1,\pi_1,\mu_1),\ldots,\ren(B_n,\pi_n,\mu_n).
  \]
  where $\textit{prop}(\cQ') =
  ((B_1,\pi_1,\mu_1),\ldots,(B_n,\pi_n,\mu_n))$.

\item Finally, for a \textsf{parallel} step
  $\tuple{(A,\pi,\mu),\cQ;memo}
  \stackrel{\textit{p}}{\longrightarrow}_\sigma
  \tuple{\cQ_1,\cQ_2,\cQ_3,\cQ_4,\cQ;memo$ $\cup\{(A,\pi,\mu)\}}$, we produce
  a residual rule of the form
  \[
  \begin{array}{llll}
  \ren(A,\pi,\mu) & \leftarrow &
  \ren(B_1,\pi_1,\mu_1),\ldots,\ren(B_n,\pi_n,\mu_n), \\
  && (\ren(B_{n+1},\pi_{n+1},\mu_{n+1}),\ldots,\ren(B_m,\pi_m,\mu_m) \\
  && \& ~\ren(B_{m+1},\pi_{m+1},\mu_{m+1}),\ldots,\ren(B_k,\pi_k,\mu_k)),\\
  && \ren(B_{k+1},\pi_{k+1},\mu_{k+1}),\ldots,\ren(B_l,\pi_l,\mu_l).
  \end{array}
  \]
  where \\
  $
  \begin{array}{lll}
    \cQ_1 = ((B_1,\pi_1,\mu_1),\ldots,(B_n,\pi_n,\mu_n)),~\\
    \cQ_2 = ((B_{n+1},\pi_{n+1},\mu_{n+1}),\ldots,(B_m,\pi_m,\mu_m)),\\
    \cQ_3 = ((B_{m+1},\pi_{m+1},\mu_{m+1}),\ldots,(B_k,\pi_k,\mu_k)),~\\
    \cQ_4 = ((B_{k+1},\pi_{k+1},\mu_{k+1}),\ldots,(B_l,\pi_l,\mu_l)).
  \end{array}
  $
\end{itemize}

\begin{example} \label{exlast}
  For instance, for the derivations of Example~\ref{ex3}, we produce
  the following residual program:
  \[
  \begin{array}{l@{~}l@{~}ll@{~~~~~~~~~~~~~~~~~}l}
    fibonacci\_{1}\_{1\_2}(0,1). \\
    fibonacci\_{1}\_{1\_2}(1,1). \\
    fibonacci\_{1}\_{1\_2}(M,N) & \leftarrow &
    M>1,~ (M1 ~is~ M-1, fibonacci\_{1}\_{1\_2}(M1,N1) \\
    && ~~~~~~~~~~~~~\& ~M2 ~is~ M-2, fibonacci\_{1}\_{1\_2}(M2,N2)), \\
    && N ~is~ N1+N2.  \\
  \end{array}
  \]
\end{example}

\subsection{Correctness and Termination Issues} 

The core of our new proposal mainly involves new control strategies,
but the main procedure is still an instance of the standard partial
evaluation framework, so its correctness should not be an issue.  In
particular, our partial evaluation scheme can be seen as an instance
of the procedure of Benkerimi and Lloyd \citeyear{BL89}, though in our
case an atom is closed only if it is a variant (rather than an
instance) of an already partially evaluated atom. Our approach is
correct though since we add calls to the predicates of the original
program for non-closed atoms (and the residual program includes a copy
of the original program clauses).

Regarding the termination of partial evaluation, this is a well
studied area and the approach that we consider based on the
homeomorphic embedding ordering is quite standard \cite{Leu02}.

Regarding the introduction of parallel conjunctions, in this paper we
assume the correctness of the underlying groundness and dependency
analyses. Moreover, we prove in the online appendix the correctness of
the few functions introduced to propagate groundness and sharing
patterns at partial evaluation time, $entry$ and $prop$. Of course,
the correctness of function $partition$ can only be ensured when
$\cQ'_2$ and $\cQ'_3$ only contain user defined predicates or ``safe''
built-ins (i.e., built-ins without side effects, which do not depend
on or may change the order of evaluation, etc).

To summarize, this paper is not concerned with the development of new
theoretical developments regarding partial evaluation or program
parallelization, but with the design of new control strategies that
could allow us to improve existing partial evaluation techniques and
use them to extract some implicit independent AND-parallelism of logic
programs. Moreover, the proof-of-concept implementation of a
parallelizing partial evaluator (that we discuss in the next section)
shows that our approach is indeed viable in practice.

\section{Experimental Evaluation} \label{experiments}

A prototype implementation of the parallelizing partial evaluator
described so far has been developed. It consists of approx.\ 1000
lines of SWI Prolog code (including the groundness call and success
pattern analysis, comments, etc).  The only missing component is the
sharing analysis, which currently should be provided by the user.  In
general, built-in's and extra-logical features are not unfolded,
though our tool includes information regarding the propagation of
groundness and sharing information for them.

A web interface to our tool is available at
\verb$http://kaz.dsic.upv.es/litep.html$.

We have tested it by running some typical benchmarks from the
literature on automatic independent AND-parallelization of logic
programs (see, e.g., \cite{MBBH99,GH09}):
\begin{itemize}
\item \textbf{amatrix} implements the addition of two matrices
  (a matrix is a list of lists);
\item \textbf{fib} computes the well-known Fibonacci function;
\item \textbf{flatten} is used to flatten a list of lists of
  any nesting depth into a flat list;
\item \textbf{hanoi} solves the Towers of Hanoi problem; 
\item \textbf{msort} implements the mergesort algorithm on lists;
\item \textbf{mmatrix} implements the multiplication of two matrices;
\item \textbf{palin} recognizes (list) palindromes;
\item \textbf{qsort} implements the quicksort algorithm on lists;
\item \textbf{tak} computes the Takeuchi function.
\end{itemize}
Moreover, in order to test the scalability of the tool, we have also
applied our parallelizing partial evaluation tool to itself
(\textbf{ppeval}). The code of the examples can be found in the tool's
webpage.

We use SWI Prolog's \textsf{concurrent/3} to run goals in
parallel. Parallel processes in SWI Prolog, however, are not
lightweight. As mentioned in \cite{SWI}, \textit{if the goals are CPU
  intensive and normally all succeeding, typically the number of CPUs
  is the optimal number of threads. Less does not use all CPUs, more
  wastes time in context switches and also uses more memory.}  For
instance, the unbound number of threads that would be created with the
program of Example~\ref{exlast} would perform very badly for even
small input values. In order to solve this problem, we replace calls
to \textsf{concurrent/3} by a special version as follows:
\begin{verbatim}
concurrent_k(A,B,C) :-
  current_threads(N), max_threads(K),!,
  (N < K -> M is N+1, 
            retractall(current_threads(_)),assert(current_threads(M)), 
            concurrent(2,[B,C],[]),
            current_threads(T), S is T-1,
            retractall(current_threads(_)),assert(current_threads(S)), 
         ;  call(A) ).
\end{verbatim}
Basically, given queries $\cQ_1$ and $\cQ_2$,
$concurrent\_k((\cQ_1,\cQ_2),\cQ'_1,\cQ'_2)$ determines, depending on
the current and maximum number of threads, if a sequential goal
$(\cQ_1,\cQ_2)$ or a parallel goal $\cQ'_1\& \cQ'_2$ should be run
(where $\cQ'_i$ is the parallel version of $\cQ_i$).

Table~\ref{times} summarizes our experimental results for the selected
benchmarks. We executed SWI-Prolog (Multi-threaded, 64 bits, Version 6.0.2)
on a 2.66 GHz Quad-Core Intel Xeon (with 8GB 1066 MHz DDR3 RAM)
running Mac OS X v10.7.3. Therefore, one can expect the best results
for a maximum of 4 threads. 
Run times have been obtained using SWI Prolog's \textsf{get\_time/1},
which is similar to SICStus \textsf{walltime} and includes CPU time,
garbage collection, etc. Rather than timings, we show the relative
speedup (i.e., run time of the original program/run time of the
residual program; values $>1$ are then actual speedups) for each
original program (column \textbf{Seq}), and its partially evaluated
version using 1/2/4/6/8 cores (columns
\textbf{Par1}/\textbf{Par2}/\textbf{Par4}/\textbf{Par6}/\textbf{Par8}).
Here, \textsf{SO} indicates a \textit{stack overflow}.

\begin{table}
\caption{Experimental evaluation of the parallelizing partial evaluator}\label{times}
\centering
$
\begin{array}{rrrrrrrrrr|}
  \hline\hline
  \mbox{\bf benchmark} & \mathbf{Seq} & \mathbf{Par1} & \mathbf{Par2}
  & \mathbf{Par4} & \mathbf{Par6} & \mathbf{Par8} \\\hline

  \mathbf{fib} & 1.00 & 1.00 & 1.83 & 2.88 & 3.82 & 3.70 \\
  \mathbf{hanoi} & 1.00 & 1.27 & 1.54 & 2.29 & 2.05 & 1.97 \\
  \mathbf{mmatrix} & 1.00 & 1.05 & 1.07 & 1.09 & 1.08 & 1.07 \\
  \mathbf{palin} & 1.00 & 1.07 & 1.79 & 2.52 & 2.30 & 2.41 \\
  \mathbf{tak} & 1.00 & 0.98 & 1.31 & 1.31 & 1.30 & 1.31 \\

  \noalign{\vspace{2ex}}

  \mathbf{amatrix} & 1.00 & 1.02 & 0.59 & 0.30 & 0.20 & 0.16 \\
  \mathbf{flatten} & 1.00 & 1.23 & 0.72 & 0.63 & 0.61 & 0.81 \\
  \mathbf{msort} & 1.00 & 1.59 & 0.86 & 1.23 & 1.22 & 1.26 \\
  \mathbf{qsort} & 1.00 & 1.73 & 0.48 & 0.71 & 0.72 & 0.60 \\

  \noalign{\vspace{2ex}}

  \mathbf{ppeval} & 1.00 & 1.00 & 1.15 & \mathsf{SO} & \mathsf{SO} & \mathsf{SO} \\
  \hline\hline
\end{array}
$\\[-3ex]
\end{table}

First, we observe that the values of column \textbf{Par1} are not
always 1.00. This is due to the effects of the partial evaluation. We
tried to minimize it, but it seems that for some examples it still
has a significant effect. The first group of benchmarks (\textbf{fib},
\textbf{hanoi}, \textbf{mmatrix}, \textbf{palin} and \textbf{tak})
show the expected results: \textbf{Par1} is generally close to 1 and
the introduction of parallel threads produces noticeable speedups.  For the
second group of benchmarks (\textbf{amatrix}, \textbf{flatten},
\textbf{msort} and \textbf{qsort}), we get a slowdown in almost all
cases but in \textbf{msort} (and, even in this case, the sequential
partial evaluation is faster). Let us take a look at the results.  For
instance, for \textbf{amatrix}, we transform:
\begin{verbatim}
  amatrix([L1|O1],[L2|O2],[L3|O3]) :- am1(L1,L2,L3), amatrix(O1,O2,O3).
\end{verbatim}
into
\begin{verbatim}
  amatrix_par([A|B],[C|D],[E|F]) :- 
                     concurrent_k((am1(A,C,E),amatrix(B,D,F)),
                                  am1(A,C,E), amatrix_par(B,D,F)).
\end{verbatim}
and leave the rest of the program untouched. For \textbf{quicksort},
we get
\begin{verbatim}
quicksort_par([A|B],C) :- partition(B,A,D,E),
                          concurrent_k((quicksort(D,F),quicksort(E,G)),
                                       quicksort_par(D,F),
                                       quicksort_par(E,G)),
                          append(F,A,G,C).
\end{verbatim}
and the rest of the program is not modified. Similar results are
obtained for \textbf{flatten} and \textbf{msort}. Note that the output
of our tool is perfectly reasonable (i.e., it coincides with a typical
parallelization by hand). So what explains the slowdowns produced?
Besides the particularities of these benchmarks, it might be caused by
the implemented model of parallel threads in SWI Prolog (which copies
ground arguments instead of sharing them). Further investigating this
point is a subject of ongoing research; e.g., we plan to test the
benchmarks using a different Prolog environment supporting
source-level primitives for AND-parallelism.
As for the third group, \textbf{ppeval}, we do not get a significant
speedup but it allows us to check that the approach is viable in
practice and scales up well to medium programs (the stack overflow
corresponds to running the specialized partial evaluator to partially
evaluate itself on 4 or more threads, and seems to be related to the
limited size of threads' stacks|i.e., it is not a fault of
\textbf{ppeval}).

In summary, the experimental evaluation is still preliminary, but it
clearly shows that there is a good potential for improving program
performance by using a parallelizing partial evaluator. Indeed, one
can easily judge by visual inspection of the annotated programs (check
the results in \verb$http://kaz.dsic.upv.es/litep.html$) that our
parallelizing partial evaluator uncovers as much parallelism
opportunities as it is possible.
%
We have not compared our tool with any existing parallelizing compiler
for logic programs yet. On the one hand, because our tool is not yet
mature enough to deal with realistic Prolog applications. On the other
hand, because we could not find a publicly available working system
for source-level program parallelization.

\section{Concluding Remarks and Future Work} \label{future}

In this work, we have presented a novel approach to parallelizing
partial evaluation. Analogously to standard approaches to automatic
independent AND-parallelization of logic programs, our partial
evaluator uses run-time groundness and dependency
information. However, in contrast to these approaches, we can
transform the source program in order to expose more opportunities for
parallelization. We are not aware of any previous proposal along the
same lines. \cite{CD92,STK97} considers performing \emph{partial
  evaluation} in parallel, which is quite a different goal as
ours. The closer approach we are aware of is that of \cite{SB94},
where a standard partial evaluator is used to expose some low level
operations of a program so that a parallelization algorithm can be
more successfully applied. They consider, however, two independent
actions: standard partial evaluation and program parallelization, in
contrast to ours. Nevertheless, the idea of combining partial
evaluation and static analysis is not new \cite{Jones97}. Also, the
use of partial evaluation to compile an instrumented interpreter can
be used to enrich source programs with some additional information
that can be useful for debugging or optimizing execution (see, e.g.,
\cite{Debois04,Jones04}). Although we are not aware of using it for
generating annotations for parallelism so far, partially evaluating an
interpreter instrumented with groundness and sharing information (so
that conjunctions are executed in parallel when safe) could get
similar results as our approach.

Being a novel approach, we consider that there is plenty of room for
further improvements. Firstly, one can consider the use of more
accurate groundness and sharing analysis. Secondly, our partition
procedure to extract two independent subgoals that can be run in
parallel is rather simple. We plan to extend it to allow an arbitrary
number of parallel subgoals, and also to allow the reordering of some
subgoals. We would also like to explore other notions of
AND-parallelism like non-strict independent AND-parallelism or, even,
dependent AND-parallelism. Finally, the combination of our approach
with a more aggressive partial evaluation scheme is also an interesting
avenue for future work.

\subsection*{Acknowledgements}

The author wants to thank the members of the SWI Prolog mailing list
for being so helpful and responsive. We also gratefully acknowledge
the anonymous referees for many useful comments and suggestions.




\includepdf[pages=-]{appendix.pdf}

\end{document}


\maketitle

\thispagestyle{myheadings}
\pagestyle{myheadings}

\begin{appendix}
\section{}

In this appendix, we state and prove the correctness of functions
$entry$ (i.e., the correctness of our notion of extended SLD
resolution) and $prop$, as defined in the body of the paper.

First, we recall the definition of function $entry$:

\begin{definition}[entry procedure] \label{entrydef} Let $H \leftarrow
  B_1,\ldots,B_n$ be a clause and $(A,\pi,\mu)$ an extended atom such
  that $A$ and $H$ unify. We denote with $entry$ a function that
  propagates $\pi$ and $\mu$ to $B_1,\ldots,B_n$. Formally,
  $entry(\pi,\mu,(H\leftarrow B_1,\ldots,B_n)) =
  ((B_1,\pi_1,\mu_1),\ldots,(B_n,\pi_n,\mu_n))$ if, for all
  $B_i=p_i(t_{i1},\ldots,t_{im_i})$, $i=1,\ldots,n$, the following
  conditions hold:
    \begin{itemize}
    \item $j \in \pi_i$ iff $\var(t_{ij})\subseteq\var(\pi(H))$ (i.e.,
      all variables in $t_{ij}$ are ground in $H$ according to $\pi$).

     \item $\{1,\ldots,m_i\}\supseteq\{j_1,\ldots,j_k\}\in\mu_i$ iff
       there are (non necessarily different) variables
      $(x_{j_1},\ldots,x_{j_k}) \in
      (\var(t_{ij_1}),\ldots,\var(t_{ij_k}))$ such that for every pair
      of different variables $x_{j_r},x_{j_s}$, we have
      $(x_{j_r},x_{j_s})\in\mu(H)$ (i.e., either the terms share some
      variable or have different variables that are shared in $H$
      according to $\mu$).
    \end{itemize}
\end{definition}
Now, let us introduce the following notion of \emph{safeness} that
will become useful to prove the correctness results.

\begin{definition}[safeness]
  Let $p(t_1,\ldots,t_n)$ be a \emph{run-time} call. We say that a
  groundness \emph{call} pattern $\pi$ is safe for $p(t_1,\ldots,t_n)$
  if $i\in\pi$ implies that $\var(t_i)=\emptyset$. Also, a sharing
  \emph{call} pattern $\mu$ is safe for $p(t_1,\ldots,t_n)$ if
  $\var(t_i)\cap\var(t_j)\neq\emptyset$ for some
  $i,j\in\{1,\ldots,n\}$, $i\neq j$, implies that $i,j\in s_i$ and
  $i,j\in s_j$, where $\mu = \tuple{s_1,\ldots,s_n}$.  This notion is
  extended to queries in the natural way.

  Analogously, let $p(t_1,\ldots,t_n)$ be a run-time call with
  \emph{computed answer substitution} $\theta$. We say that a
  groundness \emph{success} pattern $\pi'$ is safe for
  $p(t_1,\ldots,t_n)\theta$ if $i\in\pi'$ implies that
  $\var(t_i\theta)=\emptyset$. Also, a sharing \emph{success} pattern
  $\mu'$ is safe for $p(t_1,\ldots,t_n)\theta$ if
  $\var(t_i\theta)\cap\var(t_j\theta)\neq\emptyset$ for some
  $i,j\in\{1,\ldots,n\}$, $i\neq j$, implies that $i,j\in s_i$ and
  $i,j\in s_j$, where $\mu' = \tuple{s_1,\ldots,s_n}$.  This notion is
  also extended to queries in the natural way.

  Given an extended atom $(A,\pi,\mu)$ (typically with a \emph{partial
    evaluation} call $A$), we say that it is safe if, for all run-time
  call $A\theta$, both $\pi$ and $\mu$ are safe for $A\theta$.
  
  Given a partial evaluation call $A$ with call and success groundness
  (resp.\ sharing) pattern $pred(A): \pi \mapg \pi'$ (resp.\
  $pred(A):\mu\maps\mu'$), we say that this call and success pattern
  is safe for $A$ if for all run-time call $A\sigma$ and computed
  answer substitution $\theta$, the fact that $\pi$ (resp.\ $\mu$) is
  safe for $A\sigma$ implies that $\pi'$ (resp.\ $\mu'$) is safe for
  $A\sigma\theta$.
\end{definition}
Now, we recall the notion of extended SLD resolution:

\begin{definition}[extended SLD resolution]
  Extended SLD resolution, denoted by $\leadsto$, is a natural
  extension of SLD resolution over extended queries.  Formally, given
  a program $P$, an extended query $\cQ=
  (A_1,\pi_1,\mu_1),\ldots,(A_n,\pi_n,\mu_n)$, and a computation rule
  $\cR$, we say that $\leftarrow \cQ \leadsto_{P,\cR,\sigma}\:
  \leftarrow \cQ'$ is an \emph{extended SLD resolution step} for $\cQ$
  with $P$ and $\cR$ if the following conditions hold:\footnote{We
    often omit $P$, $\cR$ and/or $\sigma$ in the notation of an
    extended SLD resolution step when they are clear from the
    context.}
  \begin{itemize}
  \item $\cR(\cQ) = (A_i,\pi_i,\mu_i)$, $1\leq i\leq n$, is the selected
    extended atom,
  \item $H \leftarrow B_1,\ldots,B_m$ is a renamed apart clause of
    $P$, 
  \item $A_i$ and $H$ unify with $\sigma = mgu(A_i,H)$, and
  \item $\cQ' = entry(
    \pi_i,\mu_i,(H \leftarrow
    B_1,\ldots,B_m))\sigma$.\footnote{We let
      $((B_1,\pi_1,\mu_1),\ldots,(B_n,\pi_n,\mu_n))\sigma =
      (B_1\sigma,\pi_1,\mu_1),\ldots,(B_n\sigma,\pi_n,\mu_n)$.}
  \end{itemize}
\end{definition}
The next lemma states the correctness of the extended SLD
resolution. We only consider \emph{atomic} extended queries, which is
enough for our purposes. Moreover, information is not propagated
between query atoms; this will be the purpose of function $prop$
below.

\begin{lemma} \label{entrylemma} Let $P$ be a program and let
  $\leftarrow (A,\pi,\mu) \leadsto_{\sigma}\: \leftarrow
  (B_1\sigma,\pi_1,\mu_1),\ldots,(B_n\sigma,\pi_n,\mu_n)$ be an
  extended resolution step with clause $H\leftarrow
  B_1,\ldots,B_n$. If $(A,\pi,\mu)$ is safe, then
  $(B_i\sigma,\pi_i,\mu_i)$ is safe too, where
  $\cR((B_1\sigma,\pi_1,\mu_1),\ldots,(B_n\sigma,\pi_n,\mu_n)) =
  (B_i\sigma,\pi_i,\mu_i)$ for some selection strategy $\cR$.
\end{lemma}

\begin{proof}  
  Consider that $entry(\pi,\mu,(H\leftarrow B_1,\ldots,B_n)) =
  (B_1,\pi_1,\mu_1),\ldots,(B_n,\pi_n,\mu_n)$. 
  We prove that $(B_i\sigma,\pi_i,\mu_i)$ with $B_i =
  p_i(t_1,\ldots,t_m)$ is safe by contradiction.
  For this purpose, we consider a run-time call $A\theta$ such that
  $A\theta \hookrightarrow_\delta B_1\delta,\ldots,B_n\delta$, where
  $\delta=mgu(A\theta,H)$ (so $B_i\delta$ is an instance of
  $B_i\sigma$).

  Assume that there exists some $j\in\pi_i$ such that
  $\var(t_{j}\delta)\neq\emptyset$. Therefore, there exists some
  variable $x\in\var(t_{j})$ such that $x\delta$ is not ground. By
  Definition~\ref{entrydef}, we have that $x\in\var(\pi(H))$. However,
  since $(A,\pi,\mu)$ is safe, $\pi(A\theta)$ must be ground, and
  therefore $x\delta$ must be ground after unifying $A\theta$ with $H$
  using $\delta=mgu(A\theta,H)$, so that we get a contradiction.

  Assume now that $\mu_i = \tuple{s_1,\ldots,s_m}$ and we have
  $j,k\not\in s_j$, $j,k\not\in s_k$ but
  $\var(t_{j}\delta)\cap\var(t_{k}\delta)\neq\emptyset$.  Therefore,
  either a) there exists some variable
  $x\in\var(t_{j})\cap\var(t_{k})$ such that $x\delta$ is not ground
  or b) there are different variables
  $(x,y)\in(\var(t_{j}),\var(t_{k}))$ such that
  $\var(x\delta)\cap\var(y\delta)\neq\emptyset$.
  Consider the first case a). Here, we get immediately a contradiction
  since $j,k$ must belong to sets $s_j$ and $s_k$ by
  Definition~\ref{entrydef}. Consider now case b). Since $(A,\pi,\mu)$
  is safe and $x,y$ are bound to a term sharing variables in run-time
  call $A\theta$, we have $(x,y)\in\mu(H)$. Therefore, by
  Definition~\ref{entrydef}, $j,k$ must belong to both sets $s_j$ and
  $s_k$, and we get a contradiction too.
\end{proof}
Let us now recall the definition of function $prop$:

\begin{definition}[pattern propagation]
  Let $\cQ_1,\cQ_2$ be extended queries, with $\cQ_1 =
  (A_1,\pi_1,\mu_1),\ldots,(A_n,\pi_n,\mu_n)$ and $\cQ_2 =
  (A_{n+1},\pi_{n+1},\mu_{n+1}),\ldots,(A_m,\pi_m,\mu_m)$. We define
  the function $\mathit{prop}$ to propagate success patterns to the
  right as follows:\footnote{Note the non-standard use of function
    $entry$ to propagate success patterns to the right, despite the
    fact that $A_1\leftarrow A_2,\ldots,A_m$ is not really a program
    clause.}
  \begin{itemize}
  \item $\mathit{prop}(\cQ_1,\cQ_2) = \cQ_2$ if $n= 0$ (i.e.,
    $\cQ_1$ is an empty query);
  \item $\mathit{prop}(\cQ_1,\cQ_2) =
    ((A_1,\pi_1,\mu_1),\mathit{prop}(\cQ'_1,\cQ'_2))$ if $n>0$,\\
    $\pred(A_1):\pi_1\mapg\pi'_1$, $\pred(A_1):\mu_1\maps\mu'_1$,\\
    $\mathit{entry}(\pi'_1,\mu'_1,(A_1 \leftarrow A_2,\ldots,A_m)) =
    (A_2,\pi'_2,\mu'_2),\ldots,(A_m,\pi'_m,\mu'_m)$, \\
    $\cQ'_1 = (A_2,\pi_2\sqcap\pi'_2,\mu_2\sqcup\mu'_2),\ldots,
    (A_n,\pi_n\sqcap\pi'_n,\mu_n\sqcup\mu'_n)$, and \\ 
    $\cQ'_2 =
    (A_{n+1},\pi_{n+1}\sqcap\pi'_{n+1},\mu_{n+1}\sqcup\mu'_{n+1}),\ldots,
    (A_m,\pi_m\sqcap\pi'_m,\mu_m\sqcup\mu'_m)$.
  \end{itemize}
\end{definition}
Finally, we prove the correctness of this function:

\begin{lemma} \label{proplemma}
  Let $(A,\pi,\mu)$ be a safe extended atomic query such that
  $\leftarrow (A,\pi,\mu) \leadsto_\sigma \leftarrow \cQ$. Then
  $prop(\cQ,true)$ is a safe extended query if the considered call and
  success patterns are safe.
\end{lemma}

\begin{proof}
  We consider that $\cQ$ has two extended atoms to simplify the proof
  (the extension to arbitrary atoms can be easily done by induction on
  the number of atoms). Let $H\leftarrow B_1,B_2$ with
  $entry(\pi,\mu,(H\leftarrow B_1,B_2)) =
  (B_1,\pi_1,\mu_1),(B_2,\pi_2,\mu_2)$ and $\sigma=mgu(A,H)$, so that
  $\cQ = (B_1\sigma,\pi_1,\mu_1),(B_2\sigma,\pi_2,\mu_2)$. By
  Lemma~\ref{entrylemma}, we know that both $(B_1\sigma,\pi_1,\mu_1)$
  and $(B_2\sigma,\pi_2,\mu_2)$ are safe at clause entry (i.e., if
  selected first). Since we consider a left-to-right selection rule,
  only $(B_1\sigma,\pi_1,\mu_1)$ is safe in principle.

  Now consider the computation of
  $prop(((B_1\sigma,\pi_1,\mu_1),(B_2\sigma,\pi_2,\mu_2)),true)$. For
  this purpose, we consider the following safe call and success
  groundness and sharing patterns: $pred(B_1):\pi_1\mapg\pi'_1$ and
  $pred(B_1): \mu_1\maps\mu'_1$. Let
  $entry(\pi'_1,\mu'_1,(B_1\sigma\leftarrow B_2\sigma)) =
  (B_2\sigma,\pi'_2,\mu'_2)$. Now, we want to prove that
  \[
  prop(((B_1\sigma,\pi_1,\mu_1),(B_2\sigma,\pi_2,\mu_2)),true) =
  (B_1\sigma,\pi_1,\mu_1),(B_2\sigma,\pi_2\sqcap\pi'_2,\mu_2\sqcup\mu'_2)
  \]
  is a safe query. For this purpose, we only have to prove that
  $(B_2\sigma,\pi_2\sqcap\pi'_2,\mu_2\sqcup\mu'_2)$ is safe since
  $(B_1\sigma,\pi_1,\mu_1)$ is already proved safe under a
  left-to-right selection strategy, as mentioned above.  Let us
  consider a run-time call $B_1\sigma\theta$, together with an
  arbitrary computed answer substitution $\delta$ for
  $B_1\sigma\theta$, so that $B_2\sigma\theta\delta$ is a run-time
  call too.  We prove the claim by contradiction.

  Assume that $B_2\sigma = p(t_1,\ldots,t_n)$ and that there is some
  $i\in\pi_2\sqcap\pi'_2$ such that
  $\var(t_i\theta\delta)\neq\emptyset$. By definition, $i\in\pi_2$ or
  $i\in\pi'_2$. By Lemma~\ref{entrylemma}, we have that $\pi_2$ is
  safe at clause entry, so $i\not\in\pi_2$ since
  $\var(t_i\theta)\neq\emptyset$. By a similar argument to that of
  Lemma~\ref{entrylemma} (it again requires an application of function
  $entry$), we have that $\pi'_2$ is safe when $B_1\sigma\theta$
  succeeds, so $i\not\in\pi'_2$ too since
  $\var(t_i\theta\delta)\neq\emptyset$, and we get a contradiction.

  Consider now that
  $\var(t_j\theta\delta)\cap\var(t_k\theta\delta)\neq\emptyset$ but
  $(j,k)\not\in s_j$ and $j,k\not\in s_k$, where $\mu_2\sqcup\mu'_2 =
  \tuple{s_1,\ldots,s_n}$. Since $(B_2\sigma,\pi_2,\mu_2)$ is safe at
  clause entry, then
  $\var(t_j\theta)\cap\var(t_k\theta)=\emptyset$. Therefore, it must
  be $\delta$ that introduces some additional sharing. However, by
  applying a similar argument as that of Lemma~\ref{entrylemma}, we
  have that $\mu'_2$ is safe too when $B_2\sigma\theta$ succeeds, so
  $(j,k)\in s'_j$ and $(j,k)\in s'_k$ with $\mu'_2 =
  \tuple{s'_1,\ldots,s'_n}$ and, thus, $(j,k)\in s_j$ and $(j,k)\in
  s_k$, which gives a contradiction to our previous assumption.
\end{proof}
Finally, the correctness of function $partition$ is an easy
consequence of Lemma~\ref{proplemma}. Of course, correctness is only
ensured when $\cQ'_2$ and $\cQ'_3$ only contain user defined
predicates or ``safe'' built-ins (i.e., built-ins without side
effects, which do not depend on or may change the order of evaluation,
etc).

\end{appendix}